\documentclass[12pt]{iopart}
\usepackage{iopams}  
\usepackage{upgreek,textgreek}
\usepackage{multirow}
\usepackage{amssymb}
\usepackage{cite}
\usepackage{graphicx}
\usepackage{epsfig}
\usepackage{subfigure}
\usepackage{calc}
\usepackage{amstext}
\usepackage{multicol}
\usepackage{pslatex}

\begin{document}

\newcommand{\bfrho}{\mbox{\boldmath $\rho$}}
\newcommand{\bffeta}{\mbox{\boldmath $\eta$}}
\newcommand{\bfdelta}{\mbox{\boldmath $\delta$}}
\newcommand{\bfgamma}{\mbox{\boldmath $\gamma$}}
\newcommand{\bfsigma}{\mbox{\boldmath $\sigma$}}
\newcommand{\bftau}{\mbox{\boldmath $\tau$}}
\newcommand{\bm}{\boldmath}	
	
\bibliographystyle{iopart-num}

\title{Partially Coherent 
	 Sources with Inhomogeneous Azimuth}
	
	\author{Juan Carlos Gonz\'{a}lez de Sande$^1$, Rosario Mart\'{i}nez-Herrero$^2$, Gemma Piquero$^2$ and David Maluenda$^2$}
	\address{$^1$ ETSIS de Telecomunicaci\'{o}n, Universidad Polit\'ecnica de Madrid, Campus Sur, 28031 Madrid, Spain}
	\address{$^2$ Departamento de \'{O}ptica, Universidad Complutense de Madrid, 28040 Madrid, Spain}	
	\ead{jcgsande@ics.upm.es}
	
	\begin{abstract}
A new model of physically realizable electromagnetic source is proposed. The source is partially coherent and non-uniformly totally polarized. The coherence and polarization characteristics of this new source are analyzed. The spatial coherence area of the source can be easily modified at will. The state of polarization is linear across the transverse plane of the source with an azimuth that varies from point to point in a different way depending on the selected values of the parameters that define the source.		
	\end{abstract}
	
	%
	\vspace{2pc}


\section{Introduction}
\label{sec:introduction}
\noindent
Recently there is a great interest in proposing new sources of light with special characteristics of coherence and polarization including non uniformly polarized fields with controllable distributions across the transversal section \cite{Ambrosini:JMO94,Gori:JMO98,Seshadri:JOSAA99,Gori:OL00,Gori:JOSAA01,Gori:JOPTA01,Piquero:OC02,Paakkonen:OE02,Shirai:JOA05,RMH:OC08,RMH:OE08,Zhan:AOP09,Brown:OE10,Liang:OL14,Rodrigo:OL15,Borghi:OL15,Mei:OE16,Wang:OE16,Santarsiero:OL17,Santarsiero:OL17b}.
Works can be found in the literature that deal with partially coherent and partially or totally polarized sources, from both theoretical and experimental points of view \cite{Friberg:OC82,Friberg:JOSAA88,Serna:JMO92,Santarsiero:JOSAA99,Piquero:OC02,Santarsiero:JOSAA09,Ramirez:OC10,Sande:OE12,Maluenda:OE13}. The goal for proposing different kind of light sources generally is to get a better performance of different optical devices and techniques where specific characteristics of the used light are required depending on the application.

A big issue when a new source model is proposed, is to prove its physical realizability. The nonnegativeness condition must be satisfied by the cross spectral density function in the scalar case or by the cross spectral density matrix in the vectorial case \cite{ManWolf95}. Recently, some results have been derived that solve this task \cite{Gori:OL07,Gori:JOSAA08,RMH:OL09,RMH:OL09b,Gori:JOPTA09}. 

In this paper we introduce a new class of vectorial source whose cross spectral density matrix (CSDM) ensures its physical realizability. The source is partially coherent  non-uniformly  totally polarized and present the particularity that its state of polarization is linear with inhomogeneous azimuth. 
  
The paper is structured as follows: 
in Section 2, some concepts and parameters to be used in the present work are defined; in Section 3, the new CSDM is presented together with a proposal to experimentally synthesize it. We also characterize the source by means of different parameters as the degree of polarization, Stokes parameters, radial polarization content, and degree of coherence; finally in Section 4, the main conclusions of this work are derived.

\section{Preliminaries}
\label{Theory}
\noindent Stochastic, statistically stationary electromagnetic light sources can be appropriately described by their corresponding cross-spectral density matrix  $\widehat{W}\left( \bf{r}_1, \bf{r}_2 \right)$ \cite{ManWolf95,MHerrero:PPL09}. The vectors ${\bf{r}_j}=\left( r_j,\theta_j\right) $ with $j=1,2$ are position vectors across the source plane with $\left( r_j,\theta_j\right)$ denoting the radial and angular coordinates, respectively.

Recently, a necessary and sufficient condition has been established that guarantee that a proposed matrix represents a genuine CSDM \cite{RMH:OL09b}. This condition is an extension of the previously derived superposition rule \cite{Gori:OL07,RMH:OL09} that guarantee the nonnegativeness condition for being a genuine cross-spectral density in the scalar treatment \cite{ManWolf95}. 

In order to evaluate the coherence properties of a electromagnetic source, we will use the scalar function $g\left( \bf{r}_1,\bf{r}_2 \right)$  defined as \cite{Gori:OL07a,RMH:OL07,RMH:OL07a}
\begin{eqnarray}
\label{gr1r2}
g\left( \bf{r}_1,\bf{r}_2 \right) & = & \mu^2_{STF} 
\frac{2 \left| \det \left\lbrace \widehat{W}\left( \bf{r}_1, \bf{r}_2 \right)  \right\rbrace \right| }{ {\rm tr} \left\lbrace \widehat{W}\left( \bf{r}_1, \bf{r}_1 \right)  \right\rbrace  {\rm tr} \left\lbrace \widehat{W}\left( \bf{r}_2, \bf{r}_2 \right)  \right\rbrace}  \, 
\end{eqnarray} 
where $\mu^2_{STF}$ is the electromagnetic degree of coherence \cite{Tervo:OE03,Setala:OL04}, and "tr" and "det" denote the trace and determinant of a matrix, respectively.

From physical point of view, this quantity can be understood as the intimate capability of the field to improve their fringe visibility, under unitary transformation, in a suitable Young interference.

On the other hand, the polarization characteristics of the field can be described by its polarization matrix  resulting from the evaluation of its CSDM  at the same point $\bf{r}=\bf{r}_1= {\bf{r}}_2 $ 
\cite{ManWolf95}.

A useful tool to describe the state of polarization of the field is the Stokes vector ${\bf S}=(S_0,S_1,S_2,S_3)^T$ \cite{B&W80,Goldstein03}, where superscript $T$ denotes transpose. For non uniformly polarized beams, ${\bf S}$  is a point dependent vector and it can be derived from the polarization matrix as \cite{ManWolf95}
\begin{eqnarray}
S_i\left( \bf{r} \right)={\rm{tr}} \left\lbrace \widehat{W}\left( \bf{r}, \bf{r} \right) \widehat{\sigma}_i \right\rbrace \, , 
\end{eqnarray}
where  $\widehat{\sigma}_i$ are the $2\times 2$ identity matrix together with the three Pauli matrices~\cite{ManWolf95}
\begin{eqnarray}
\begin{array}{c c}
	\widehat{\sigma}_0=\left( \begin{array}{c c}
	1&0 \\0&1\\
	\end{array}\right) \, , \,\,\, \,\,\, &
	\widehat{\sigma}_1=\left( \begin{array}{c c}
	1&0 \\0& -1\\
	\end{array}\right) \, , \\
	\widehat{\sigma}_2=\left( \begin{array}{c c}
	0&1 \\1&0\\
	\end{array}\right) \, , \,\,\, \,\,\, &
	\widehat{\sigma}_3=\left( \begin{array}{c c}
	0&-{\rm {i}} \\{\rm {i}}&0\\
	\end{array}\right) \, .
\end{array}
\end{eqnarray}

An important parameter describing the polarization properties of the field is the degree of polarization, $P$, that gives the ratio of the polarized part of the field to the total irradiance \cite{ManWolf95}. It can be obtained from the polarization matrix as
\begin{equation}
P=\sqrt{1-\frac{4 \det \left\lbrace \widehat{W}\left( \bf{r}, \bf{r} \right)  \right\rbrace}{\left( {\rm tr} \left\lbrace \widehat{W}\left( \bf{r}, \bf{r} \right)  \right\rbrace \right)^2}} \, .
\end{equation}

For linearly polarized light the azimuth, $\psi\left( \bf{r} \right)$, can be related to the Stokes parameters by \cite{B&W80,Goldstein03}
\begin{equation}
\tan 2 \psi\left( \bf{r} \right)=\frac{S_2\left( \bf{r} \right)}{S_1\left( \bf{r} \right)} \, .
\end{equation}

In order to get an insight of the  polarization characteristics of this source, global parameters \cite{RMH:OC08}  as the radial and azimuthal polarized content of the field can be analyzed. The irradiance ratio of the radial (or azimuthal) component of the field to the total irradiance are related to the Stokes parameters of the field as \cite{RMH:OC08}
\begin{equation}
\rho_R \left( \bf{r} \right)=\frac{1}{2}+\frac{\cos 2 \theta}{2}\frac{S_1\left( \bf{r} \right)}{S_0\left( \bf{r} \right)}+\frac{\sin 2 \theta}{2}\frac{S_2\left( \bf{r} \right)}{S_0\left( \bf{r} \right)} \, 
\end{equation} 
and
\begin{equation}
\rho_{\theta}\left( \bf{r} \right)=\frac{1}{2}-\frac{\cos 2 \theta}{2}\frac{S_1\left( \bf{r} \right)}{S_0\left( \bf{r} \right)}-\frac{\sin 2 \theta}{2}\frac{S_2\left( \bf{r} \right)}{S_0\left( \bf{r} \right)} \, .
\end{equation} 
Note that $0 \le \rho_R \left( \bf{r} \right) \le 1$ ($0 \le \rho_{\theta}\left( \bf{r} \right) \le 1$), 0 meaning that the radial (azimuthal) component is null and 1 that the field is radially (azimuthally) polarized. It can be observed that $\rho_R \left( \bf{r} \right)+\rho_{\theta}\left( \bf{r} \right)=1$ so they carry complementary information. They are point dependent parameters for non uniformly polarized fields. Their average over the region where the field irradiance is significant can be obtained as \cite{RMH:OC08}
\begin{equation}
\label{rhoR}
\widetilde{\rho}_R =\frac{\int\rho_R \left( \bf{r} \right) {\rm{tr}} \left\lbrace \widehat{W}\left( \bf{r}, \bf{r} \right) \right\rbrace d \bf{r} }{\int {\rm{tr}} \left\lbrace \widehat{W}\left( \bf{r}, \bf{r} \right) \right\rbrace d \bf{r}}  \, ,
\end{equation} 
and
\begin{equation}
\widetilde{\rho}_{\theta} =\frac{\int\rho_{\theta} \left( \bf{r} \right) {\rm{tr}} \left\lbrace \widehat{W}\left( \bf{r}, \bf{r} \right) \right\rbrace d \bf{r} }{\int {\rm{tr}} \left\lbrace \widehat{W}\left( \bf{r}, \bf{r} \right) \right\rbrace d \bf{r}}  \, .
\end{equation} 

\section{Source Model and Characteristics}
\label{CSDM}

\noindent  In the present work, we introduce an new class of electromagnetic sources whose CSDM is given by 
\begin{equation}
\widehat{W}\left(  \bf{r}_1, \bf{r}_2 \right)=\widehat{V}^{\ast}\left( \bf{r}_1 \right)  \widehat{W}_{0}\left( \bf{r}_1, \bf{r}_2 \right) \widehat{V}\left( \bf{r}_2 \right) \, .
\label{W}
\end{equation}
being $\widehat{V}\left( \bf{r} \right)$  the following deterministic matrix 
\begin{eqnarray}
\widehat{V}\left( \bf{r} \right)=\left( \begin{array}{c c}
f_m(r) \cos \beta_m  & 0 \\
0 & f_n(r) \cos \beta_n  \\
\end{array}\right) \, ,
\label{Vmatrix}
\end{eqnarray}
where $m$ and $n$ are integers, $\beta_m=m\theta-\alpha$ with  $\alpha$ an arbitrary constant angle, and $f_m(r)$ are real functions of the radial coordinate that must be zero at the origin in order to avoid singularities. On the other hand,  $\widehat{W}_{0}\left( \bf{r}_1, \bf{r}_2 \right)$ matrix is of the form
\begin{equation}
\widehat{W}_{0}\left(\bf{r}_1, \bf{r}_2 \right) =\tau^{\ast}\left( \bf{r}_1\right) \tau\left( \bf{r}_2\right) G\left( \bf{r}_1-\bf{r}_2  \right) \widehat{A} \, ,
\label{W0}
\end{equation}
where  the function $\tau\left( \bf{r} \right)$ is  an arbitrary complex function and $\widehat{A}$ is a constant $2 \times 2$ matrix that satisfy $\det \left\lbrace  \widehat{A} \right\rbrace =0$ so the proposed source is totally polarized. Finally, the function $G\left( \bf{r}_1- \bf{r}_2 \right)$ is defined as
\begin{equation}
G\left( \bf{r}_1- \bf{r}_2 \right)=\int h^2\left( \brho \right) \exp \left[ {\rm i} k  \brho \cdot \left( \bf{r}_1- \bf{r}_2 \right)  \right]  d \bfrho
\label{gFT}
\end{equation}
being $h\left( {\brho} \right)$ a real function and $k$ the wavenumber. 

From a physical point of view, $\widehat{W}_{0}\left( \bf{r}_1, \bf{r}_2 \right)$ can be interpreted as the CSDM of a partially coherent Schell type source. 

It can be proven that the matrix given in Eq. (\ref{W}) with the chosen $\widehat{V}\left( \bf{r} \right)$ and $\widehat{W}_{0}\left(\bf{r}_1, \bf{r}_2 \right)$ (Eqs. (\ref{Vmatrix}) and (\ref{W0})) is a genuine CSDM \cite{Gori:OL07,RMH:OL09,RMH:OL09b}. Then, the proposed source described by Eq. (\ref{W}) is physically realizable.
 
\begin{figure*}
  	\centering
  	\includegraphics[width=0.8\linewidth]{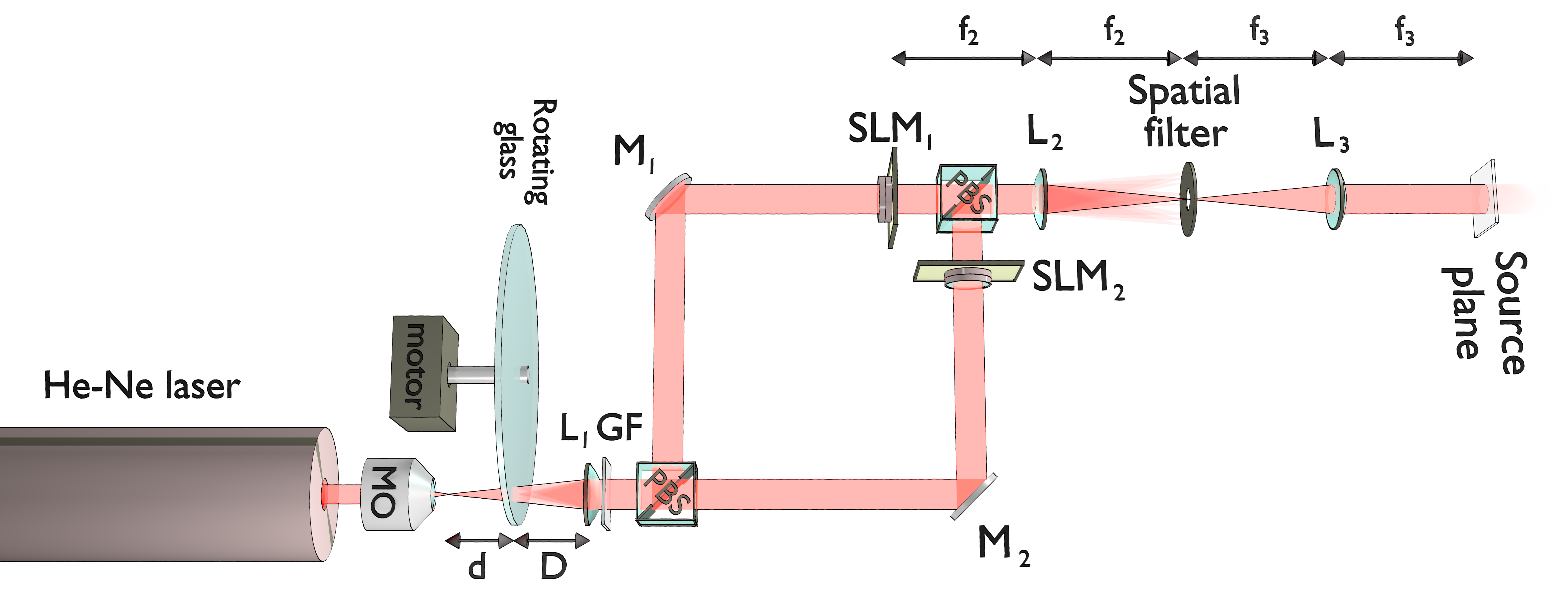}
  	\caption{Experimental set-up for obtaining the proposed source.}
  	\label{Fig_Setup}
  \end{figure*}

 Figure \ref{Fig_Setup} shows a schematics of a proposal for an experimental set-up in order to obtain the proposed source. A TEM$_{00}$ He-Ne laser linearly polarized  at $\pi/4$ is expanded by microscope objective (MO) and, after a distance $d$, a rotating ground glass generates an incoherent beam with Gaussian irradiance profile and transverse  width $w_0$. Note that the size of the incoherent source can be tuned by changing the distance $d$. After freely propagating a distance $D$, from the van Cittert-Zernike theorem follows that the transverse coherent length is \cite{MHerrero:PPL09}
  \begin{equation}
  \label{mu}
  \mu = \frac{D}{\sqrt{2}}\,\frac{\lambda}{\pi w_0} \, ,
  \end{equation}
where $\lambda$ is the wavelength, thus, we can control the degree of coherence of the source on varying the distance $D$. Afterwards, the L$_1$ lens with focal length $D$ collimates the incoherent beam and a Gaussian filter (GF) performs its irradiance profile.

 In order to synthesize the non uniformly polarized field, we transform the source given by Eq. (\ref{W0}) by means of a set-up based on a modified Mach-Zehnder interferometer where $x-$ and $y-$component of the beam are independently manipulated by a spatial light modulator (SLM) in each arm of the interferometer \cite{Maluenda:OE13}. The complex transmittance introduced with the SLM can be described by a point dependent $2\times 2$ matrix, in this case this matrix is $\widehat{V}\left( \bf{r} \right)$, so the resulting CSDM given by Eq. (\ref{W}) is implemented in the source plane.
 
To characterize the coherence properties, we analyze the behavior of parameter $g\left( \bf{r}_1,\bf{r}_2 \right)$ given by Eq. (\ref{gr1r2}). For our source 
this parameter reads
\begin{equation}
g\left( \bf{r}_1,\bf{r}_2 \right)=\frac{\left| G\left( \bf{r}_1-\bf{r}_2 \right)\right|^2}{\left| G\left( 0 \right)\right|^2 } \, .
\end{equation} 

Then, we have that $g\left( \bf{r}_1,\bf{r}_2 \right)$ is independent of $m$, $n$ and $\alpha$ parameters. However, it can be controlled by modifying the transverse coherence length of the source that impinges on the modified Mach-Zender interferometer (see figure \ref{Fig_Setup} and Eq. (\ref{mu})).

In this work, we consider the matrix $\widehat{A}$ corresponding to a linearly polarized field at $\pi/4$, that is
\begin{eqnarray}
\widehat{A}=a_0 \left( \begin{array}{c c}
1 & 1 \\ 1 & 1 \\
\end{array}\right) \, .
\label{Amtrix}
\end{eqnarray}
In this case, the following polarization matrix is obtained
\begin{eqnarray}
\label{Pol}
\widehat{W}\left( \bf{r}, \bf{r} \right)&=&\widehat{V}^{\ast}\left( \bf{r} \right)  \widehat{W}_0\left( \bf{r}, \bf{r} \right) \widehat{V}\left( \bf{r} \right) 
\left|\tau\left( \bf{r} \right)\right|^2 G(0) \widehat{V}^{\ast}\left( \bf{r} \right) \widehat{A}\;  \widehat{V}\left( \bf{r} \right) .
\end{eqnarray}
 
The resulting source is partially coherent and totally polarized  because $\det \widehat{W}\left( \bf{r}, \bf{r} \right)=0$. However the state of polarization changes from  point to point in a way determined by the spatial dependence of the matrix $\widehat{V}\left( \bf{r} \right)$. 
In fact, evaluation of Eq. (\ref{Pol}) results in the following polarization matrix 
\begin{eqnarray}
\widehat{W}\left( \bf{r}, \bf{r} \right)&=& a_0\left|\tau\left( \bf{r} \right)\right|^2 G(0)  
\left( \begin{array}{c c}
p_{xx}\left( r,\theta\right)  & p_{xy}\left( r,\theta\right)  \\
p_{yx}\left( r,\theta\right)  & p_{yy}\left( r,\theta\right) \\
\end{array}\right) \, ,
\label{W1}
\end{eqnarray}
where the  values for the $p_{ij}$ elements are
\begin{equation}
p_{xx}\left( r,\theta\right) =f_m^2(r) \cos^2 \beta_m \, ,
\end{equation}
\begin{equation}
p_{xy}\left( r,\theta\right) =f_m(r)f_n(r)\cos \beta_m \sin  \beta_n \, ,
\end{equation}
\begin{equation}
p_{yx}\left( r,\theta\right)=p_{xy}\left( r,\theta\right) .
\end{equation}
and 
\begin{equation}
p_{yy}\left( r,\theta\right) =f_n^2(r) \sin^2  \beta_n \, .
\end{equation}

The Stokes parameters result
\begin{eqnarray}
S_0\left( r,\theta\right)&=&a_0\left|\tau\left( \bf{r} \right)\right|^2 G(0) 
\left[ f_m^2(r) \cos^2 \beta_m +f_n^2(r) \sin^2 \beta_n \right]   \, ,
\label{S0_v2}
\end{eqnarray}
\begin{eqnarray}
S_1\left( r,\theta\right)&=&a_0\left|\tau\left( \bf{r} \right)\right|^2 G(0) 
\left[ f_m^2(r) \cos^2 \beta_m - f_n^2(r) \sin^2 \beta_n \right]   \, ,
\label{S1_v2}
\end{eqnarray}
\begin{eqnarray}
S_2\left( r,\theta\right)&=&a_0\left|\tau\left( \bf{r} \right)\right|^2 G(0) 
2 f_m(r) f_n(r) \cos \beta_m  \sin \beta_n    \, ,
\label{S2_v2}
\end{eqnarray}
and 
\begin{eqnarray}
S_3\left( r,\theta\right) =0    \, ,
\label{S3_v2}
\end{eqnarray}
that is a linearly polarized field at any point of the source cross section ($S_3\left( r,\theta\right) =0 $).

The azimuth results
\begin{equation}
\tan 2 \psi\left( \bf{r} \right)=\frac{2 f_m(r) f_n(r) \cos \beta_m  \sin \beta_n }{f_m^2(r) \cos^2 \beta_m - f_n^2(r) \sin^2 \beta_n} \, .
\end{equation}
Note that this $\psi\left( \bf{r} \right)$ is a point dependent function that varies in a different way depending on the source parameters. 

For the particular selection of $m=n$, the Stokes parameters take the simple form 
\begin{eqnarray}
S_0\left( r,\theta\right)=a_0\left|\tau\left( \bf{r} \right)\right|^2 G(0)  f_m^2(r)  \, ,
\label{S0_v2}
\end{eqnarray}
\begin{eqnarray}
S_1\left( r,\theta\right)=a_0\left|\tau\left( \bf{r} \right)\right|^2 G(0)  f_m^2(r) \cos 2 \beta_m    \, ,
\label{S1_v2}
\end{eqnarray}
\begin{eqnarray}
S_2\left( r,\theta\right)=a_0\left|\tau\left( \bf{r} \right)\right|^2 G(0) f_m^2(r) \sin 2 \beta_m    \, ,
\label{S2_v2}
\end{eqnarray}
and 
\begin{eqnarray}
S_3\left( r,\theta\right) =0    \, .
\label{S3_v2}
\end{eqnarray}

In this case the azimuth only depends on the angular coordinate as $\psi\left( \theta \right)=\beta_m=m\theta-\alpha$, i.e., the azimuth rotates continuously at different rates depending on $m$ value.

The resulting radial polarization content only depends on the azimuth $\theta$ as
\begin{equation}
\rho_R \left( \theta \right)=\frac{1}{2}\left[ 1+\cos \left( 2 \theta -2 \beta_m \right) \right] \, .
\end{equation} 

When averaging this value according to Eq. (\ref{rhoR}) the following value is obtained
\begin{equation}
\widetilde{\rho}_R =\frac{1+\delta_{m,1} \cos 2 \alpha }{2} \, ,
\end{equation}  
being $\delta_{i,j}$ the Kronecker delta function. Then, the field shows a spirally polarized pattern for $m=n=1$ \cite{Gori:JOSAA01,Paakkonen:OE02}, i. e., spirally polarized beams are a particular case included in the proposed class of sources.  

It is known that spirally polarized beams show a polarization map where the azimuth rotates at the same rate than the angular coordinate $\theta$ and in the same sense (counterclockwise).

\begin{figure}[!h]
	\vspace{-0.2cm}
	\centering
	\includegraphics[width=0.4\linewidth]{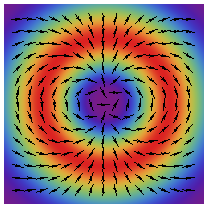}
	\caption{Polarization map for $m=n=2$ and $\alpha =pi/2$. Colors indicate irradiance level: red for high irradiance and violet for low irradiance. Arrows indicate the electric field} 
	\label{Fig_Pol_m2}
	\vspace{-0.1cm}
\end{figure}

\begin{figure}[!h]
	\vspace{-0.2cm}
	\centering
	\includegraphics[width=0.4\linewidth]{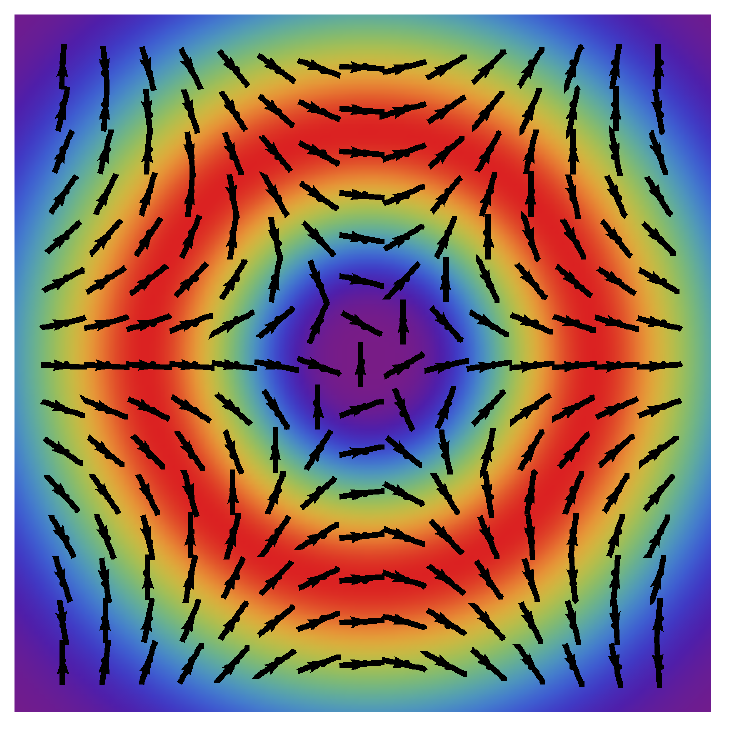}
	\caption{Same that figure \ref{Fig_Pol_m2} for $m=n=-2$.} 
	\label{Fig_Pol_m_1}
	\vspace{-0.1cm}
\end{figure}

In the more general case $m=n\ne 1$, the average of radial and azimuthal content are equal to $1/2$, independently of the parameter $\alpha$. Figure \ref{Fig_Pol_m2} shows the polarization map obtained for $m=n=2$ and $\alpha=pi/2$. For drawing this figure and the all figures below, the family of functions  $f_m(r)=br^{\left| m\right| }$ has been chosen. It can be observed that  the azimuth of the linearly polarized states rotates twice faster than for spirally polarized beams (see Figure \ref{Fig_Pol_m2}). In general,  the rotation rate of the azimuth is $|m|$ times faster than in the case of spirally polarized beams. The sense of rotation is counterclockwise for positive $m$ values and clockwise for negative $m$ values. For example, figure \ref{Fig_Pol_m_1} shows the case $m=n=-2$.

\section{\uppercase{Conclusions}}
\label{sec:conclusion}

\noindent A new class of electromagnetic sources is proposed. They present the property of being partially coherent non uniformly totally polarized sources. Once the parameters of the source have been selected, this source can be experimentally generated by means of a modified Mach-Zender interferometer where two spatial light modulators control the polarization properties of each interferometer arm.  The coherence area of the source can be easily modified by simply varying a distance in the proposed experimental setup. The state of polarization is always linearly polarized and its azimuth varies in different ways depending on the chosen values for the characteristic parameters defining the source. 
In the particular case of selecting $m=n$, for any concentric ring to the source axis, the azimuth of the polarized light rotates periodically in the whole circle and the number of periods correspond to $2|m|$ in a complete circle. The sense of rotation changes with the sign of the $m$ value.     

\section*{\uppercase{Acknowledgements}}

\noindent This work has been supported by Spanish Ministerio de Econom\'ia y Competitividad under projects FIS2013-46475 and FIS2016-75147.

\vfill
\providecommand{\newblock}{}

\vfill
\end{document}